\documentstyle[11pt,newpasp,twoside,epsf]{article}
\begin{document}
\title{Volume-limited SDSS/FIRST quasars and the radio dichotomy}
\author{S.\ Jester} \affil{Fermilab, PO Box 500, Batavia, IL 60510}
\author{R.\ G.\ Kron} \affil{Fermilab, and The University of Chicago,
Department of Astronomy and Astrophysics, 5640 South Ellis Avenue,
Chicago, IL 60637}

\begin{abstract}
Much evidence has been presented in favor of and against the existence
of two distinct populations of quasars, radio-loud and
radio-quiet. The SDSS differs from earlier optically selected quasar
surveys in the large number of quasars and the targeting of FIRST
radio source counterparts as quasar candidates. This allows a
qualitatively different approach of constructing a series of samples
at different redshifts which are volume-limited with respect to both
radio and optical luminosity. This technique avoids any biases from
the strong evolution of quasar counts with redshift and potential
redshift-dependent selection effects.  We find that optical and radio
luminosities of quasars detected in both SDSS and FIRST are not well
correlated within each redshift shell, although the
fraction of radio detections among optically selected quasars remains
roughly constant at 10\% for $z \leq 3.2$. The distribution in the
luminosity-luminosity plane does not appear to be strongly
bimodal. The optical luminosity function is marginally flatter at
higher radio luminosities.
\end{abstract}

\section{Introduction}\label{s:intro}

Quasars were first found as optical identifications of luminous radio
sources. However, only about 10\% of optically identified quasars were
radio-luminous, leading to a division of quasars into radio-``loud''
and radio-``quiet'' objects. Radio observations of optically selected
quasars (Strittmatter et al. 1980 and papers citing them) found
a bimodal distribution of the radio flux, radio luminosity, or ratio
of radio to optical flux. This has resulted in many claims that there
are two distinct populations, although it is still unclear whether
they should be distinguished by considering the radio luminosity or
the radio-optical flux ratio, and what the physical origin of this
bimodality or dichotomy might be. More recently, the FIRST radio
survey has filled gaps in the radio-to-optical flux ratio distribution
found in earlier, shallower surveys (White et al 2000), suggesting
that the radio-loud and radio-quiet objects are instead the extremes
of a continuum of sources.

The SDSS quasar survey targets both optically selected quasar
candidates and point-like optical counterparts of FIRST
sources. Together with the large number of sources compared to
previous surveys, this allows to construct a series of volume-limited
samples. This avoids any potential biases arising from the rapid
evolution of quasar counts and luminosities with redshift. Therefore,
the SDSS allows to take a different look at the bimodality
question. Once selection effects are taken into account, this approach
amounts to constructing the bivariate radio-optical quasar luminosity
function.

\section{Results}\label{s:results}

We use data (including rest-frame $i$-band luminosities) from the
quasar catalog (Schneider et al. 2003; also Schneider et al. in this
volume) created from the SDSS Data Release~1 (Abazajian et al. 2003).
We restrict ourselves to FIRST-selected quasar candidates and those
identified by the ``low-redshift UV excess'' ($ugri$) part of the
color selection algorithm (Richards et al. 2002). This algorithm
selects targets with non-stellar colors down to $i =19.1$ and has been
sufficiently uniform and complete for DR1 data so that first results
can be obtained without any corrections.

We construct a series of volume-limited samples of radial extent
$\Delta z = 0.2$, retaining only objects with luminosities between
that corresponding to the faint flux limit at the upper redshift
boundary ($i = 19.1$ for SDSS, $f_{\mathrm{1.4\,GHz}} =
1\,\mathrm{mJy}$ for FIRST) and that corresponding to the bright limit
at the lower redshift boundary ($i=15$ for SDSS; FIRST practically has
no bright limit).

\subsection{Relation between optical and radio luminosity}
\begin{figure}
\plotfiddle{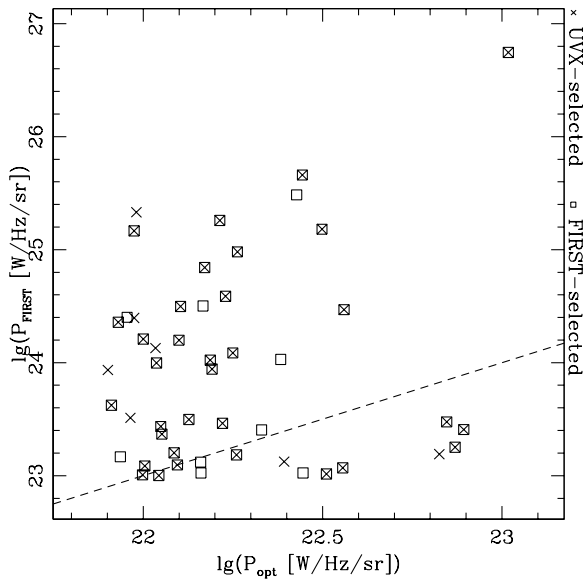}{0.5\textwidth}{0}{40}{40}{-120}{-70}
\caption{\label{f:LLex} Example of distribution of quasars in the
  Luminosity-Luminosity plane for $0.4 \le z < 0.6$. X symbols are
  color-selected (UV excess) quasars, open boxes are FIRST-selected
  targets. The dashed line shows the traditional division between
  radio-loud and radio-quiet objects at a luminosity ratio of
  $R=10$.}
\end{figure}
We construct volume-limited quasar samples between redshifts of 0.2
and 3.2. This retains a few hundred objects in each redshift shell out
of originally 10,000. Figure~\ref{f:LLex} shows an example of the
distribution of quasars making the luminosity cut of both surveys in
the luminosity-luminosity plane.  They straddle the dividing line
between radio-loud and radio-quiet objects as defined by a
radio-optical luminosity ratio of 10 (dashed line), with many objects
lying close to the line.  A caveat in the interpretation of this
result is that two simplifications have been made: the $K$-correction
did not take into account the actual spectral index of the quasar, but
assumed a spectral shape $f_\nu \propto \nu^{-0.5}$ for all objects at
all wavelengths (compare the contribution about quasar bimodality by
Ivezi\'{c} et al. in these proceedings), and the radio sources include
both flat- and steep-spectrum sources, while in principle we need to
remove intrinsically fainter flat-spectrum sources which have been
beamed above the survey limit. These corrections will be made in
future work.

In this particular case, the most luminous radio source is also the
most luminous optical source, but this is not generally true - the
maximum optical and radio luminosity in each bin are not well
correlated.  Similarly, the radio and optical luminosities of objects
within each bin are not strongly correlated, with correlation
coefficients of the logarithmic luminosities typically in the range
0-0.5.  This rules out any strict power-law dependence between radio
and optical luminosity with a scatter smaller than the luminosity
range in each redshift shell (which spans, however, only one decade or
so in optical luminosity).  Instead, it requires a hidden parameter
governing the radio luminosity relative to the optical.

In addition to the objects shown in Fig.\,\ref{f:LLex}, each redshift
shell also contains quasars in the same optical luminosity range but
which have a FIRST luminosity below the cutoff.  It is possible to
compare the radio properties of all SDSS-detected quasars by
constructing optical luminosity functions cumulated up to different
maximum radio luminosities. The optical luminosity function becomes
steeper as the maximum radio luminosity limit is lowered from the
brightest detected source to the luminosity cutoff, i.e., a luminous
radio source is more likely (but not \emph{required}) to be a luminous
optical source as well.

\subsection{Radio-loud fraction as function of redshift}

\begin{figure}
\plotfiddle{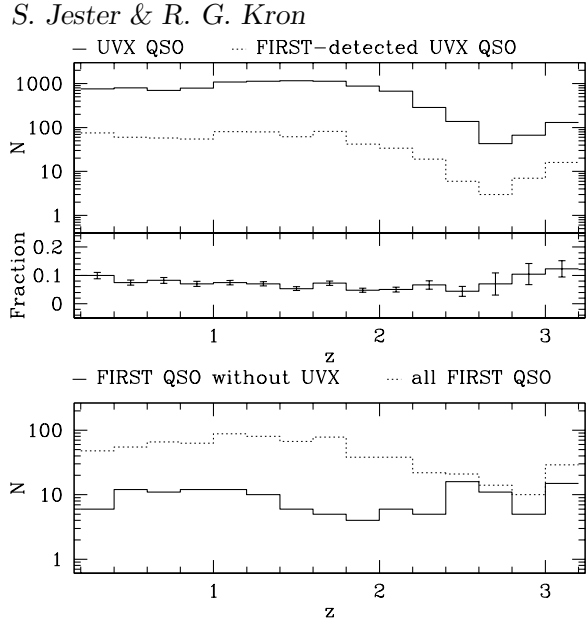}{0.5\textwidth}{0}{40}{40}{-120}{-70}
\caption{\label{f:redhist}\emph{Upper panel:} number of color-selected
quasars (now simply flux-limited; solid line) and of FIRST detections
among them (dashed line). \emph{Lower panel:} number of all quasars
targeted as optically unresolved FIRST matches (dashed line), and
number targeted \emph{only} as FIRST match, but not as color-selected
candidates (solid line).}
\end{figure}

In view of the strong evolution of quasar counts with redshift, it is
of considerable interest whether the fraction of radio-loud objects
changes with redshift.  (The magnitude of the fraction depends on the
ratio of the radio and optical flux limits and cannot in itself be
interpreted easily.)  In the case of SDSS, it is not sufficient to
consider merely the fraction of FIRST detections among all quasars in
each volume slice without completeness corrections because the survey
employs sparse sampling of the color-selected candidates in certain
redshift ranges (Richards et al. 2002), while the number of quasars
targeted as point-source identification of FIRST sources remains
unchanged (lower panel in Fig.\,\ref{f:redhist}). Ignoring this would
lead to the erroneous conclusion that the fraction of radio-detected
quasars increases drastically beyond $z=2$.

Instead, it is useful to consider the fraction of color-selected
quasars (both resolved and point sources) which have a FIRST detection
(upper panel in Fig.\,\ref{f:redhist}).  This fraction is remarkably
constant, implying a similar evolution of radio and optical luminosity
with redshift.  This finding is in apparent contradiction to the
preceding section, where we found no good radio-optical correlation on
an object-by-object basis.  The reconciliation lies in a radio-optical
correlation with a scatter larger than the luminosity range we have
considered in each redshift shell.  It remains to investigate these
correlations in more detail and to compare them to radio-optical
correlations such as those found in the most powerful radio sources
(see, e.g., Willott et al. 1999).


We are grateful to \v{Z}eljko Ivezi\'{c} for challenging discussions, to
Matt Jarvis for valuable suggestions, and to the SDSS Quasar Working
Group which has laid the foundation for this work. Funding for the
Sloan Digital Sky Survey (SDSS) has been provided by the Alfred
P. Sloan Foundation, the Participating Institutions, the National
Aeronautics and Space Administration, the National Science Foundation,
the U.S. Department of Energy, the Japanese Monbukagakusho, and the
Max Planck Society. The SDSS Web site is http://www.sdss.org/.

\end{document}